\documentclass{aastex631}

\usepackage{color}
\usepackage{hyperref}
\usepackage{epstopdf}
\usepackage{graphicx}
\usepackage[FIGTOPCAP]{subfigure}
\usepackage{CJKutf8}
\usepackage{amsmath}
\usepackage{longtable}

\begin{document}
\title{GRB~231115A: a nearby Magnetar Giant Flare or a cosmic Short Gamma-Ray Burst?}

\correspondingauthor{Yun Wang}
\email{wangyun@pmo.ac.cn}

\author[0000-0002-8385-7848]{Yun Wang}
\affiliation{Key Laboratory of Dark Matter and Space Astronomy, Purple Mountain Observatory, Chinese Academy of Sciences, Nanjing 210034, China}

\author[0000-0002-9775-2692]{Yu-Jia Wei}
\affiliation{Key Laboratory of Dark Matter and Space Astronomy, Purple Mountain Observatory, Chinese Academy of Sciences, Nanjing 210034, China}
\affiliation{School of Astronomy and Space Science, University of Science and Technology of China, Hefei, Anhui 230026, China}

\author[0000-0003-2915-7434]{Hao Zhou}
\affiliation{Key Laboratory of Dark Matter and Space Astronomy, Purple Mountain Observatory, Chinese Academy of Sciences, Nanjing 210034, China}
\affiliation{School of Astronomy and Space Science, University of Science and Technology of China, Hefei, Anhui 230026, China}

\author[0000-0002-9037-8642]{Jia Ren}
\affiliation{School of Astronomy and Space Science, Nanjing University, Nanjing 210093, China}
\affiliation{Key Laboratory of Modern Astronomy and Astrophysics (Nanjing University), Ministry of Education, China}

\author[0000-0003-4963-7275]{Zi-Qing Xia}
\affiliation{Key Laboratory of Dark Matter and Space Astronomy, Purple Mountain Observatory, Chinese Academy of Sciences, Nanjing 210034, China} 

\author[0000-0003-4977-9724]{Zhi-Ping Jin}
\affiliation{Key Laboratory of Dark Matter and Space Astronomy, Purple Mountain Observatory, Chinese Academy of Sciences, Nanjing 210034, China}
\affiliation{School of Astronomy and Space Science, University of Science and Technology of China, Hefei, Anhui 230026, China}


\begin{abstract}
There are two classes of gamma-ray transients with a duration shorter than 2 seconds. One consists of cosmic short Gamma-Ray Bursts (GRBs) taking place in the deep universe via the neutron star mergers, and the other is the magnetar giant flares (GFs) with energies of $\sim 10^{44}-10^{46}$~erg from ``nearby" galaxies. Though the magnetar GFs and the short GRBs have rather similar temporal and spectral properties, their energies ($E_{\rm \gamma,iso}$) are different by quite a few orders of magnitude and hence can be distinguished supposing the host galaxies have been robustly identified. The newly observed GRB~231115A has been widely discussed as a new GF event for its high probability of being associated with M82. Here we conduct a detailed analysis of its prompt emission observed by Fermi-GBM, and compare the parameters with existing observations. The prompt gamma-ray emission properties of GRB~231115A, if associated with M82, nicely follow the $E_{\rm p,z}$--$E_{\gamma,\rm iso}$ relation of the GFs, where $E_{\rm p,z}$ is the peak energy of the gamma-ray spectrum after the redshift ($z$) correction. To be a short GRB, the reshift needs to be $\sim 1$. Though such a chance is low, the available X-ray/GeV observation upper limits are not stringent enough to further rule out this possibility. We have also discussed the prospect of convincingly establishing the magnetar origin of GRB~231115A-like events in the future. 
\end{abstract}

\section{Introduction} \label{sec:intro}
Gamma-ray Bursts (GRBs) have been observed for over half a century, and can be classified into short GRBs and long GRBs based on their bimodal duration distribution \citep{1993ApJ...413L.101K}. Long GRBs typically originate in star-forming regions within galaxies and are observed in association with massive star-collapse supernovae \citep{1993ApJ...405..273W,2006Natur.441..463F}, while short GRBs are found in low star-forming regions of their host galaxies and are believed to originate from compact binaries \citep{1996bboe.book..682E,1992ApJ...395L..83N,2005Natur.437..851G,2010ApJ...708....9F,2010ApJ...725.1202L,2013ApJ...776...18F,2014ARA&A..52...43B}.
In addition, there are notable distinct events called giant flares (GFs) emitting from magnetars, which exhibit characteristics similar to those of short GRBs and are mixed in with the current observations \citep{2001AIPC..586..495D,2005Natur.434.1098H,2005MNRAS.362L...8L,2006MNRAS.365..885P}. When the burst is associated with a Soft Gamma-ray Repeater (SGR) or when the host galaxy is identified, several events are revealed. These include GRBs 790305 (SGR 0526-66; \citealt{1979Natur.282..587M, Mazets1982}), 980618 (SGR 1627-41; \citealt{1999ApJ...519L.151M, 1999ApJ...519L.139W, 2001ApJS..137..227A}), 980827 (SGR 1900+14; \citealt{1999Natur.397...41H, 1999AstL...25..635M, 2007ApJ...665L..55T}), and 041227 (SGR 1806-20; \citealt{2005Natur.434.1098H, 2005Natur.434.1107P, 2007AstL...33....1F}), which are associated with SGRs. Additionally, GRBs 051103 (M81; \citealt{2007AstL...33...19F}), 070201 (M31; \citealt{Mazets2008}), 070222 (M83; \citealt{2021ApJ...907L..28B}), and 200415A (NGC 253; \citealt{2021AAS...23723302R, 2021Natur.589..211S, 2020ApJ...899..106Y, 2020ApJ...903L..32Z}) are likely associated with specific galaxies.

The common characteristics of observed GFs include the rapidly rising exponential decay light curves, the hard and rapidly evolving energy spectra, and the high energies of $10^{44}-10^{46}$ erg~\citep{1999Natur.397...41H,1999AstL...25..635M,2007AstL...33....1F}. 
GRB 980618, the most intense flare observed from the SGR~1626-41, shows a distinct characteristic which is a relatively slow-rise light curve~\citep{1999ApJ...519L.151M}. However, what can be considered as a ``smoking gun" evidence is the presence of long periodic tails after initial pulses~\citep{1979Natur.282..587M, 1999Natur.397...41H, 2005Natur.434.1098H}, originating from the modulation of the rotation period of the neutron star~\citep{1983A&A...126..400B, 2005ApJ...628L..53I, 2005ApJ...632L.111S, 2006ApJ...637L.117W}. When GFs occur in other galaxies, it may only be possible to detect the initial pulses without the presence of periodic tails. Therefore, the identification of such events is both interesting and challenging.

{In this work, we conduct a detailed data analysis on the new GF candidate observed by Fermi-GBM~\citep{2023GCN.35044....1D,2023GCN.35038....1B,2023GCN.35059....1M,2023GCN.35062....1F}, and discuss the potential of observing such events in the future.} Our work is organized as follows.
In Section \ref{sec:Obs}, we provide a summary of current observations and conduct a detailed analysis of Fermi-GBM data, along with the observational upper limit of Fermi-LAT. In Section \ref{sec:ch}, we present various characteristics of the prompt emission and compare them with known GRBs and GFs. In Section \ref{sec:mo}, {we model the afterglow in both GRB and GF scenarios, search the parameter space based on existing observational bounds, and make a rough estimate on the magnetar age.} Finally, in Section \ref{sec:sd}, we summarize all features and discuss potential opportunities for known facility observations.

\section{Observersion and data analysis} \label{sec:Obs}
\begin{figure}[!htp]
    \centering
    \includegraphics[width=0.52\textwidth]{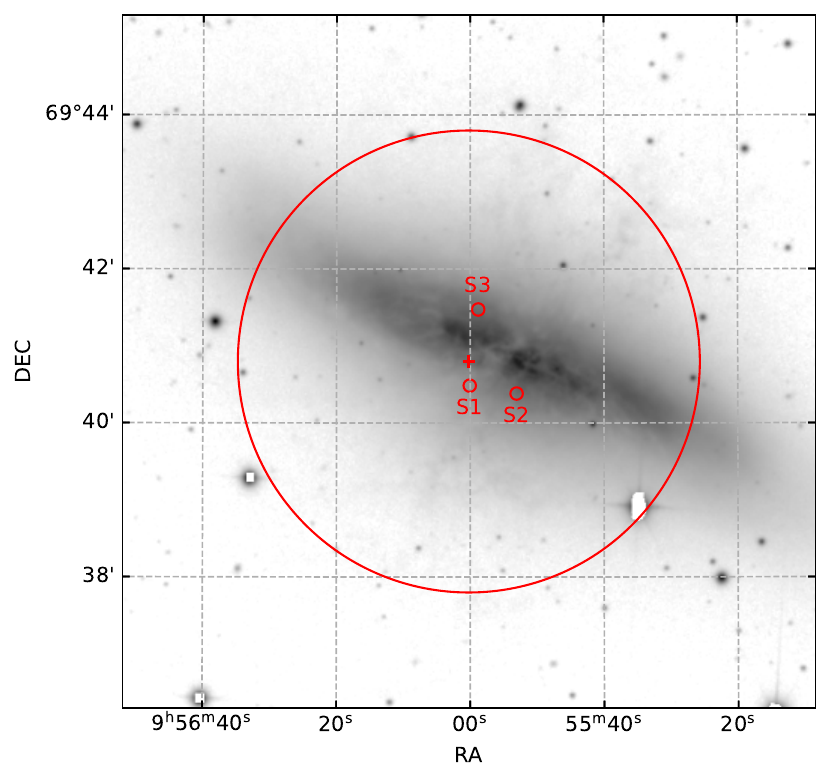}
    \caption{The ZTF r-band image of M82 field. The large red circle with a ``+" at its center shows the localization from INTEGRAL \citep{2023GCN.35037....1M}. Three small red circles marked with S1, S2 and S3 represent the position of possible transient AT2023xvj \citep{2023GCN.35041....1K}, W20231115a and W20231115b \citep{Hu_GCN}.}
    \label{fig:1}
\end{figure}

At 15:36:21 UT on November 15, 2023, GRB~231115A triggered Fermi-GBM. The position coordinates were RA~=~131.0, Dec~=~73.5 with the statistical uncertainty 8~degrees \citep{GCN35035}.
INTEGRAL \citep{2023GCN.35037....1M} have reported more accurate position uncertainty (2 arc minutes) and is consistent with the M82 galaxy (coincidence probability $\sim$ 0.03\%), marked with red ``+" (RA~=~149.0, Dec~=~69.7) in Figure~\ref{fig:1}. There have no exact multi-messenger counterparts including radio \citep{2023GCN.35070....1C}, optical \citep{2023GCN.35041....1K,Hu_GCN}, X-ray \citep{2023GCN.35054....1K,2023GCN.35064....1O,2023GCN.35175}, neutrino \citep{2023GCN.35053....1I} and gravitational waves \citep{2023GCN.35049....1L}.  
The positions of possible optical transient sources are marked with small red circles as shown in Figure~\ref{fig:1}. 

\begin{figure}[!htp]
    \centering
    \includegraphics[width=0.5\textwidth]{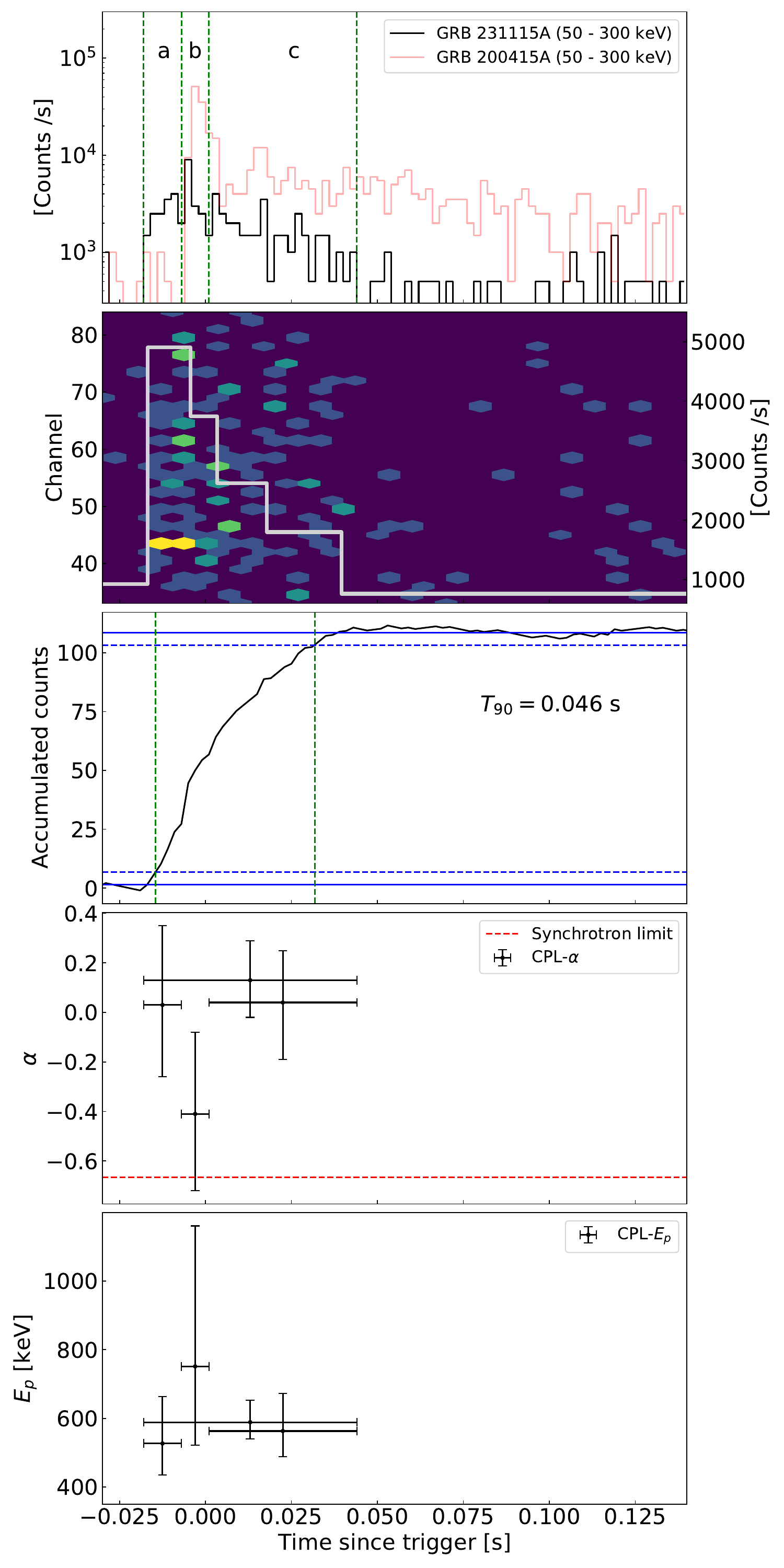}
    \caption{Observation data of Fermi-GBM and parameter evolution of CPL model. The first panel shows the light curves of GRB 231115A and GRB 200415A in the 50-300 keV energy range, and the time period used for spectral analysis. The second panel shows the count corresponding to channel {and the time bin intervals provided by the Bayesian block method.} The third panel is the photon count accumulation curve and duration from 5\% to 95\%. The two bottom panels are the evolution of the spectral index $\alpha$ and energy peak $E_p$ of the CPL model, where the red dotted line represents the synchrotron limit.}
    \label{fig:LC_PAR}
\end{figure}

Therefore, we mainly focus on what information the Fermi-GBM observations can bring to us, and further data analysis is completed by {\tt HEtools} \citep{2023ApJ...953L...8W}. The instrumentation of the Fermi-GBM payload encompasses two types of detectors: specifically, 12 sodium iodide (NaI) detectors and 2 bismuth germanate (BGO) detectors \citep{meegan2009fermi}. 
Based on the angle between the source location and the pointing of each detector, we selected a NaI (n7) detector and a BGO (b1) detector respectively.
For this extremely short duration event, we rebin Time-Tagged Event (TTE) data to obtain sufficient time resolution and custom time intervals.
{The comparison of light curves for the prompt emission from TTE data (50--300 keV) of GRB 231115A and GRB 200415A is shown in the top panel of Figure \ref{fig:LC_PAR}. The four time intervals for spectral analysis are {rising period} (a: $T_0 + [-0.018,~-0.007]~\rm s$), {peak period} (b: $T_0 + [-0.007,~0.001]~\rm s$), {decay period} (c: $T_0 + [0.001,~0.044]~\rm s$), and the whole burst ($T_0 + [-0.018,~0.044]~\rm s$). The second panel shows the energy channels recorded for TTE data and the rebin light curve (silver line) by Bayesian block method \citep{scargle2013studies}. In the third panel, we calculate $T_{90}$ ($\sim$ 46 ms) for 50--300 keV energy range based on the accumulated percentage of photon events~\citep{koshut1996systematic}.} 

\subsection{Spectral Analysis of Fermi-GBM}\label{sec:spec_ana}
As preparation for spectral analysis, we used \texttt{GBM Data Tools} \citep{GbmDataTools} to extract the spectrum files, background files, and response files of this source at the above four time intervals.
The forward fitting method in spectral analysis requires the form of the photon number spectrum to be set in advance. Here we consider two types of photon number spectrum models. The first is the cutoff power-law (CPL) model commonly used to describe GRB spectra, in the form of
\begin{equation}
    { N(E)=A\left(\frac{E}{100\,{\rm keV}}\right)^{\alpha}\exp\left(-\frac{E}{E_{\rm c}}\right)},
\end{equation}
where \emph{$\alpha$} is the power law photon spectral index, \emph{E$_{\rm c}$} is the {break} energy in the spectrum,
and the peak energy $E_{\rm p}$ is equal to $E_{\rm c}(2+\alpha)$.
Since the spectrum of GF is usually quasi thermal, another type of model is the single blackbody (BB) model and the multicolor blackbody (mBB) model \citep{ryde2010identification}. 
The BB model is expressed as
\begin{equation}
    { N(E)=\frac{ 8.0525\times K E^2}{(kT)^4 (e^{E/kT}-1)}},
\end{equation}
where \emph{kT} is the blackbody temperature keV.
Assuming that the luminosity dependence on temperature has a power law form \citep{hou2018multicolor}, the mBB model is expressed as
\begin{equation}
    N(E)=\frac{8.0525(m+1)K}{\left[\left(\frac{T_{\max}}{T_{\min}}\right)^{m+1}-1\right]}\left(\frac{kT_{\min}}{\rm keV}\right)^{-2}I(E),\label{N(E)}
\end{equation}
where
\begin{equation}
    I(E)=\left(\frac{E}{kT_{\rm min}}\right)^{m-1}\int_{\frac{E}{kT_{\rm max}}}^{\frac{E}{kT_{\rm min}}}\frac{x^{2-m}}{e^x-1}dx,\label{I(E)}
\end{equation}
where $x=E/kT$, the temperature range from $kT_{\rm min}$ to $kT_{\rm max}$, and the index $m$ of the temperature determines the shape of spectra. 

\begin{figure}[!htp]
    \centering
    \includegraphics[width=0.49\textwidth]{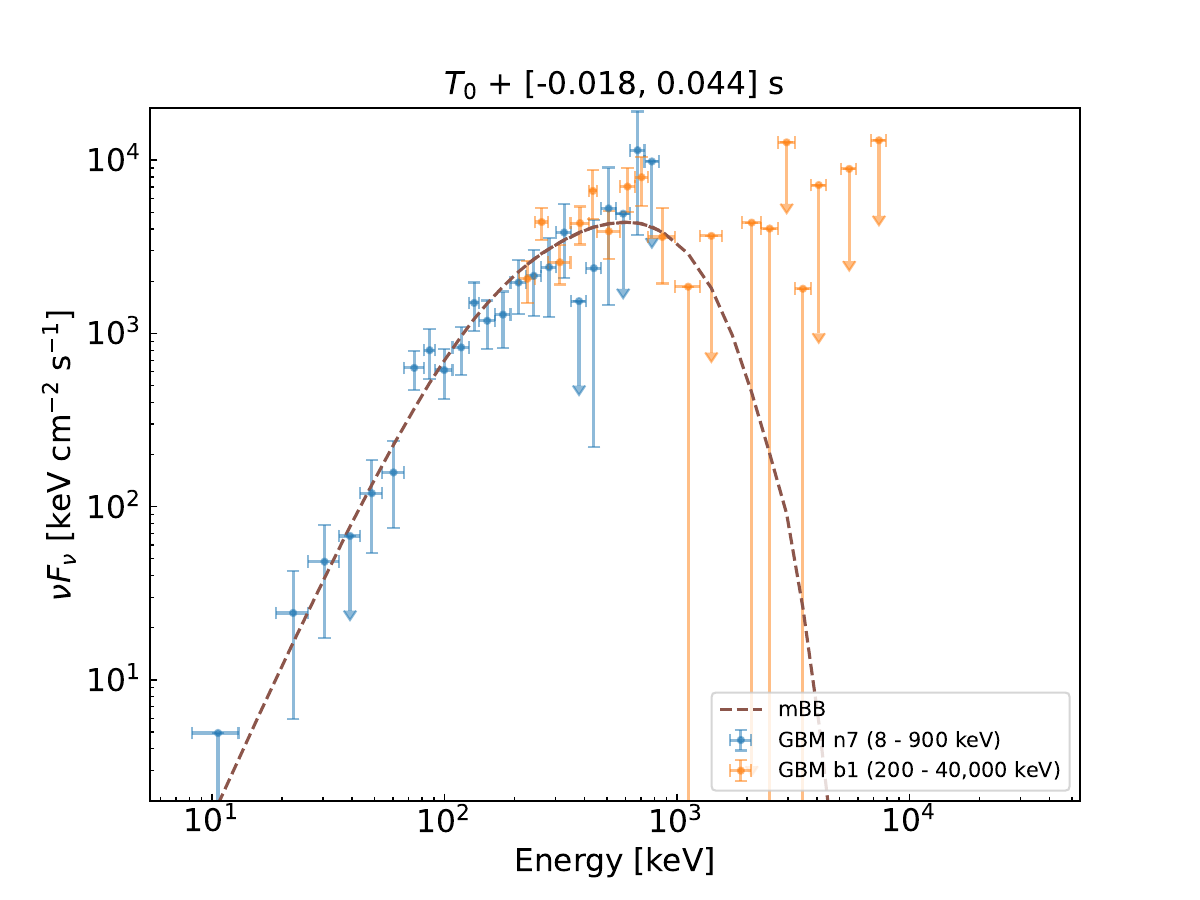}
    \includegraphics[width=0.35\textwidth]{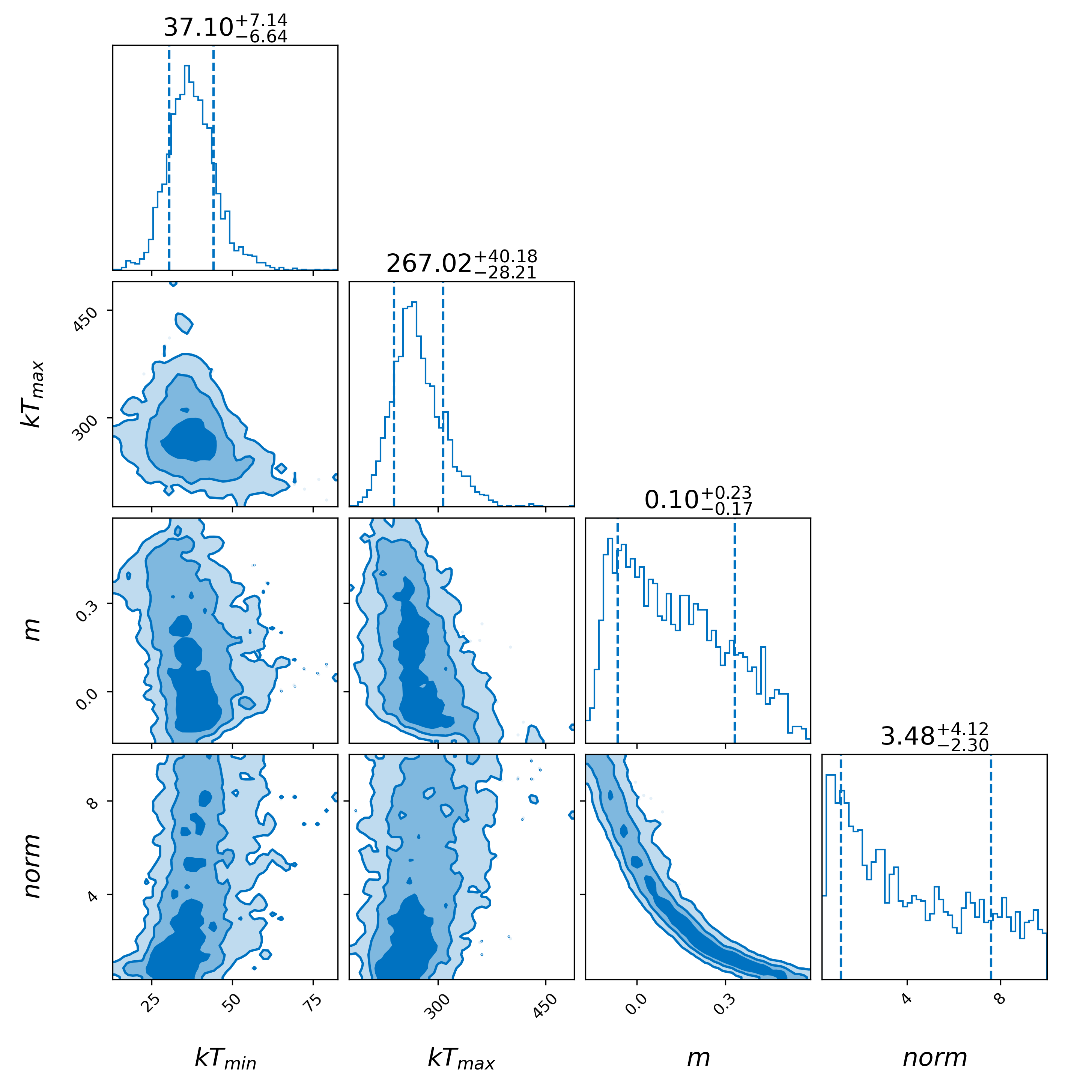}
    \caption{Time-integrated spectra fitting result. Left panel: The $\nu F_{\nu}$ spectra of GRB~231115A at integrated period. Right panel: Posterior contours of mBB model parameters.}
    \label{fig:mbb_spec}
\end{figure}

To obtain the best fit, it is common to maximize the likelihood function, which depends on the model fold response versus the observed data. For GBM data, the likelihood function used is {\tt pgstat}, refer to the {\tt XSPEC} manual\footnote{\url{https:// heasarc.gsfc.nasa.gov/xanadu/xspec/manual/XSappendixStatistics.html}}. For parameter estimation and comparison of photon spectrum models, the Bayesian inference \citep{thrane2019introduction,van2021bayesian} is widely used for this purpose. In our analysis, {\tt PyMultiNest} \citep{2014A&A...564A.125B} in the {\tt Bilby} \citep{ashton2019bilby} package serves as a sampler for Bayesian inference.
The intention of nest sampling is to provide evidence of the model, and the model selection can be done by comparing Bayes factors (BF). The BF is the ratio of the Bayesian evidence ($\mathcal{Z} = \int \mathcal{L}(d|\theta) \pi(\theta) d\theta$) for different models. 
The log of Bayes factor can be written as $\ln\text{BF}^\text{A}_\text{B} = \ln({\cal Z}_\text{A}) - \ln({\cal Z}_\text{B})$.
When $\ln{\rm BF} > 8$, we can say that there is a "strong evidence" in favor of one hypothesis over the other \citep{thrane2019introduction}.	
In addition to providing Bayesian evidence, nested sampling can also provide posterior parameter distributions. The posterior parameters of each model and the model selection results are shown in Table \ref{tab:tab1}.
In Figure \ref{fig:LC_PAR}, the two bottom panels show the evolution over time of the power law photon spectral index ($\alpha$) and the energy peak ($E_p$) respectively. It is worth noting that $\alpha$ of the CPL model exceeds the synchrotron limit, also known as the ``Line of Death" \citep{preece1998synchrotron,preece2002consistency}.
Moreover, its spectrum is harder than that of ordinary GRBs, and all intervals favor the mBB model. These characteristics indicate that the spectrum of GRB~231115A is quasi thermal, as shown in Figure \ref{fig:mbb_spec}.

\setlength{\tabcolsep}{2mm}{}
\begin{deluxetable}{ccccccccccccccccccc}
\label{tab:tab1}
\tabletypesize{\small}
\tablecaption{Spectral fitting result}
\tablehead{\colhead{Time interval} & \colhead{Model}& \colhead{$\alpha$/$m$} &\colhead{$E_{p}$/$kT$/$kT_{\rm min}$,$kT_{\rm {max}}$}  & {Flux (1-10,000 keV)}&$\ln{\cal Z}$ & {Highest evidence}\\
\colhead{$T_0$+ (s)} &\colhead{}& \colhead{} &\colhead{(keV)} &\colhead{($10^{-5}$ erg $\rm cm^{-2}$ $\rm s^{-1}$)} &\colhead{} &\colhead{} &\colhead{}}
\startdata
Time-integrated spectra \\
{[-0.018, 0.044]} & CPL & 0.16$_{-0.16}^{+0.15}$ & 579.01$_{-38.89}^{+48.35}$ & 1.36$_{-0.09}^{+0.10}$ &-219.99 \\ 
 & BB & ... & 124.00$_{-6.15}^{+6.79}$ & 1.24$_{-0.08}^{+0.08}$ & -222.11 \\ 
 & mBB & 0.10$_{-0.17}^{+0.23}$ & 37.10$_{-6.64}^{+7.14}$, 267.02$_{-28.21}^{+40.18}$ & 1.37$_{-0.10}^{+0.11}$ & -211.98 &$\checkmark^*$\\ 
\hline
Time-resolved spectra \\
{[-0.018, -0.007]} & CPL & -0.20$_{-0.13}^{+0.10}$ & 601.16$_{-71.47}^{+85.53}$ & 2.31$_{-0.26}^{+0.30}$ &-164.27 \\ 
 & BB & ... & 118.55$_{-9.02}^{+10.71}$ & 2.13$_{-0.21}^{+0.25}$ & -157.99 \\ 
 & mBB & 0.20$_{-0.15}^{+0.22}$ & 37.02$_{-11.22}^{+18.88}$, 228.99$_{-38.70}^{+48.00}$ & 2.21$_{-0.23}^{+0.28}$ & -156.29 & $\checkmark$\\ 
\hline 
{[-0.007, 0.001]} & CPL & -0.66$_{-0.12}^{+0.10}$ & 1221.93$_{-257.11}^{+371.04}$ & 4.13$_{-0.69}^{+0.91}$ &-164.79 \\ 
 & BB & ... & 151.90$_{-14.87}^{+16.59}$ & 2.97$_{-0.37}^{+0.40}$ & -166.25 \\ 
 & mBB & 0.11$_{-0.11}^{+0.17}$ & 23.12$_{-7.67}^{+11.39}$, 405.50$_{-81.23}^{+100.64}$ & 3.46$_{-0.50}^{+0.54}$ & -160.92 & $\checkmark$\\ 
\hline 
{[0.001, 0.044]} & CPL & -0.28$_{-0.08}^{+0.06}$ & 691.80$_{-71.07}^{+106.47}$ & 0.99$_{-0.10}^{+0.13}$ &-192.64 \\ 
 & BB & ... & 118.36$_{-8.88}^{+11.10}$ & 0.81$_{-0.08}^{+0.09}$ & -183.19 \\ 
 & mBB & -0.04$_{-0.11}^{+0.16}$ & 45.82$_{-8.20}^{+12.70}$, 249.29$_{-46.84}^{+59.17}$ & 0.91$_{-0.10}^{+0.11}$ & -179.21 & $\checkmark$\\ 
\hline 
\enddata
\tablecomments{The ``..." represents missing parameters of different models. ``$\checkmark^*$" represents strong evidence, that is, $\ln$BF $>$ 8.}
\end{deluxetable}

\subsection{Upper limit of Fermi-LAT}\label{sec:fermi-lat}
Fermi-LAT is a high performance gamma-ray telescope for the photon energy range from 30 MeV to 1 TeV~\citep{2021AjelloApJS}.
GRB 231115A has been found in the field of view for Fermi-LAT from ${T_0}$ to ${T_0}+3000$ s.
We select the photon events in the first 3000 s within the energy range from 500 MeV to 5 GeV with the {\tt SOURCE} event class and the {\tt FRONT+BACK} type.
In order to reduce the contamination from the Earth’s limb, we exclude photons with zenith angles larger than 100$^{\circ}$. Then we extract good time intervals with the quality-filter cut {\tt (DATA\_QUAL==1 \&\& LAT\_CONFIG==1)}.

In this work, we use the standard unbinned likelihood analysis within a 10$^{\circ}$ region of interest (ROI) centered on the INTEGRAL location of GRB 231115A. 
The initial model for the ROI region, generated by the {\tt make4FGLxml.py} script\footnote{\url{http://fermi.gsfc.nasa.gov/ssc/data/analysis/user/}}, includes the galactic diffuse emission template ({\tt gll\_iem\_v07.fits}), the isotropic diffuse spectral model for the {\tt SOURCE} data ({\tt iso\_P8R3\_SOURCE\_V3\_v1.txt}) and all the Fourth Fermi-LAT source catalog~\citep[{\tt gll\_psc\_v31.fit};][]{2020AbdollahiApJS} sources within 20 degrees around GRB 231115A. 
{The version of the Instrument Response Function (IRF) used in the analysis is {\tt P8R3\_SOURCE\_V3}.}
We model the gamma-ray emission from the GRB~231115A region as a point source at the INTEGRAL location of (ra, dec) = (149.03$^{\circ}$, 69.69$^{\circ}$) and set its spectral shape to the {\tt PowerLaw} (${\frac{dN}{dE}=N_0(\frac{E}{E_0})^\gamma}$) model with the index of $\gamma=-2$.
Due to the very short time interval we select, only normalizations of the galactic, isotropic diffuse models and spectra of GRB 231115A are set to be free parameters in the fitting process. 
Other parameters are set as the default of 4FGL.
The {\tt Fermitools} package\footnote{\url{https://github.com/fermi-lat/Fermitools-conda/}} are used in our work.
No significant gamma-ray emission from GRB 231115A was detected by Fermi-LAT.
Then a 95$\%$ confidence level upper limit of flux was derived as $1.89 \times 10^{-10}\,{\rm erg\, cm^{-2}\,s^{-1}}$ in the energy range of 0.5--5 GeV.

\section{CHARACTERISTICS} \label{sec:ch}
\subsection{Power Density Spectrum} \label{sec:PDS}
\begin{figure}[!h]
    \centering
    \includegraphics[width=0.5\textwidth]{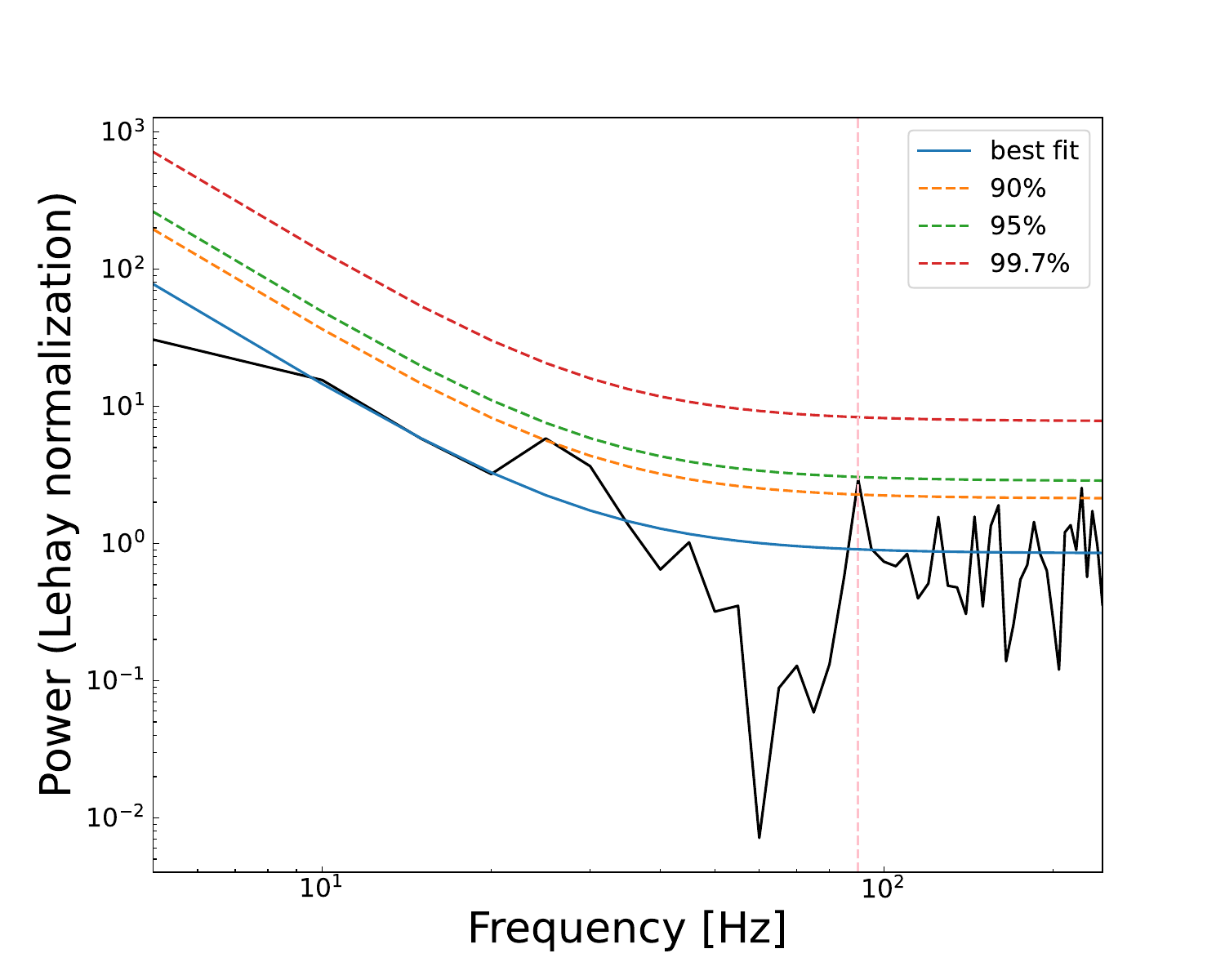}
    \caption{The Power Density Spectra (PDS). The blue solid line represents the best fit of the PDS, while the orange, green, and red dashed lines correspond to significance levels of 90\%, 95\%, and 99.7\%, respectively. The pink dashed line corresponds to the peak of the PDS occurring at $\sim$90 Hz.}
    \label{fig:PDS}
\end{figure}

{The presence of long periodic tails after initial pulses could be the strongest evidence for GF (or neutron star) origin, and the timescale of the period is about a few seconds~\citep{1979Natur.282..587M, 1999Natur.397...41H, 2005Natur.434.1098H}. In addition, short GRBs associated with binary neutron star mergers may display kilohertz quasiperiodic oscillations~\citep{2023Natur.613..253C}.}
We therefore tried to evaluate the possibility of a periodic signal in this event through the Power Density Spectrum (PDS). One of the important steps is to model red noise to calculate the significance of periodic frequencies.
For the above goal, we used a procedure based on \cite{vaughan2005simple,vaughan2010bayesian,vaughan2013random} and \cite{covino2019gamma}.
Also, we refer to \cite{beloborodov1998self,guidorzi2012average,guidorzi2016individual,dichiara2013average,dichiara2013search} for details on analyzing power density spectra in GRBs.
PDS is derived by discrete Fourier transformation and normalized according to \cite{leahy1983searches}, which fit by a single power-law function plus white noise \citep{guidorzi2016individual}, 
\begin{equation}
    S_{\rm PL}(f) = N\,f^{-\alpha} + B,
    \label{}
\end{equation}
where $N$ is a normalization factor, $f$ is the sampling frequency, and its lower limit is related to the length of the time series, which is $1/T$ (the time interval).
The upper limit of $f$ is the Nyquist frequency, which is $1/(2{\delta_t})$, where $\delta_t$ is the time bin size of data.
The value of white noise $B$ is expected to be 2, which is the expected value of a $\chi^2_2$ distribution for pure Poissonian variance in the Leahy normalization. We employ Bayesian inference for abrove model parameters by using {\tt emcee} \citep{2013PASP..125..306F}.
The maximum likelihood function we use is called \textit{Whittle} likelihood function \citep{vaughan2010bayesian}.
When enough samples are obtained, we calculate the global significance of every frequency in the PDS according to $T_{\rm R} = \max_j
R_j$, where $R = 2P/S$, $P$ is the simulated or observed PDS, and $S$ is the best-fit PDS model. 
This method selects the maximum deviation from the continuum for each simulated PDS.
The observed $T_{\rm R}$ values are compared to the simulated distribution and significance is assessed directly.
The corrections for the multiple trials performed were included in the analysis because the same procedure was applied to the simulated as well as to the real data.
As shown in Figure~\ref{fig:PDS}, the PDS exhibits a peak at $\sim$90 Hz, but it is not significant enough to be considered a quasi-periodic signal. {Perhaps due to the long distance, the periodic signals have not been identified in GRB 051103 and GRB 200415A~\citep{2020ApJ...899..106Y}, which are more powerful than GRB~231115A by an order of magnitude, either.}

\begin{figure}[!htp]
    \centering
    \includegraphics[width=0.49\textwidth]{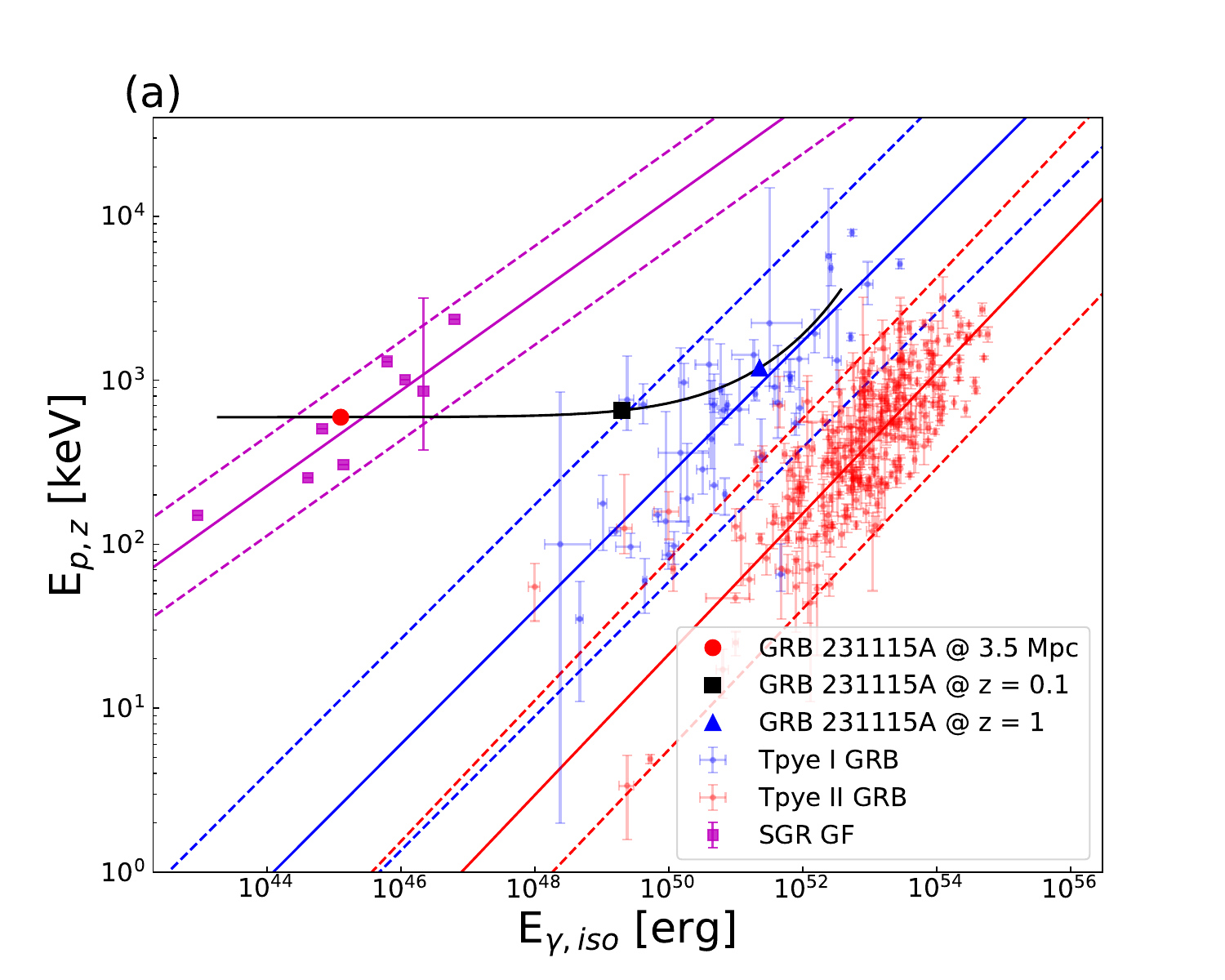}
    \includegraphics[width=0.49\textwidth]{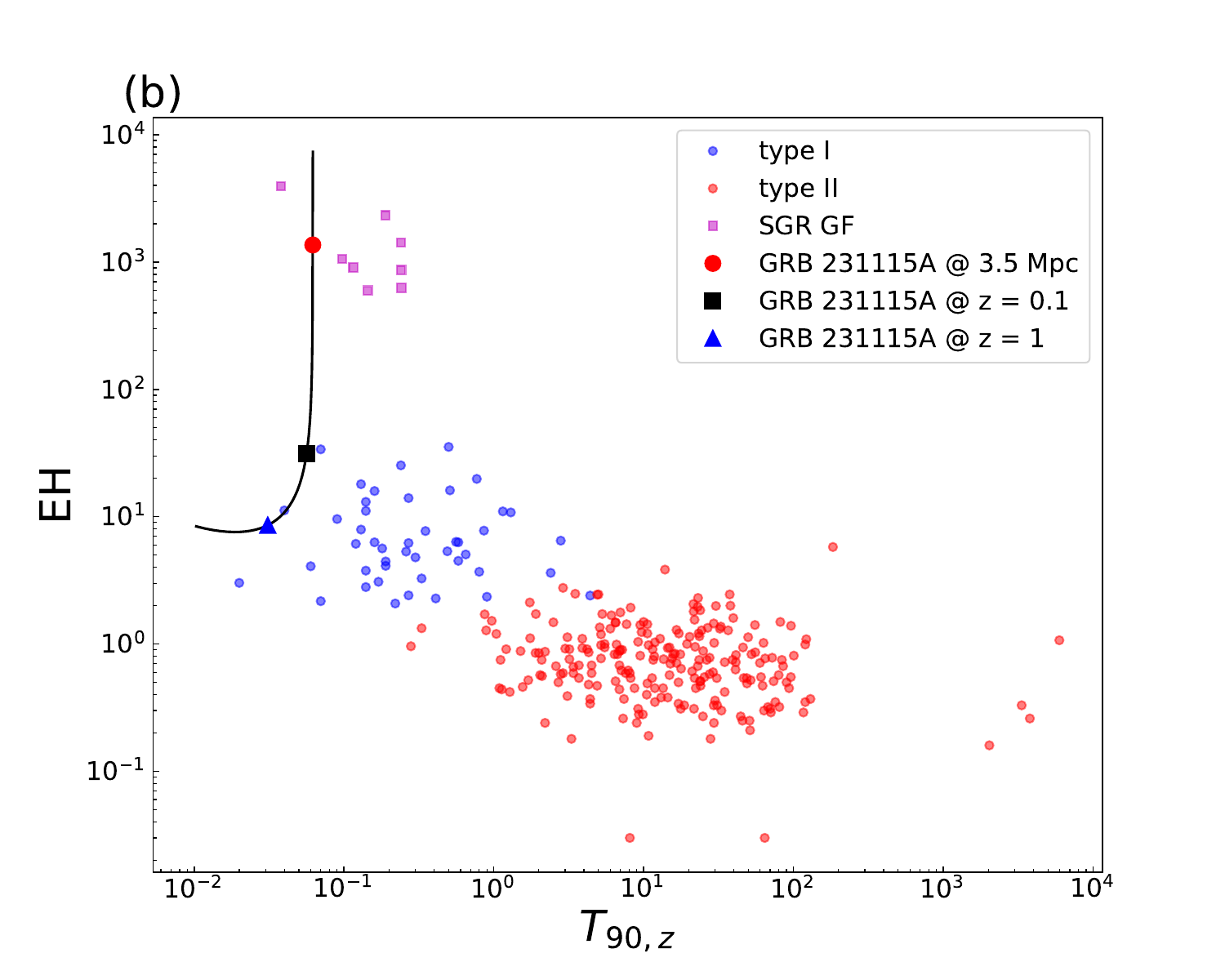}
    \includegraphics[width=0.49\textwidth]{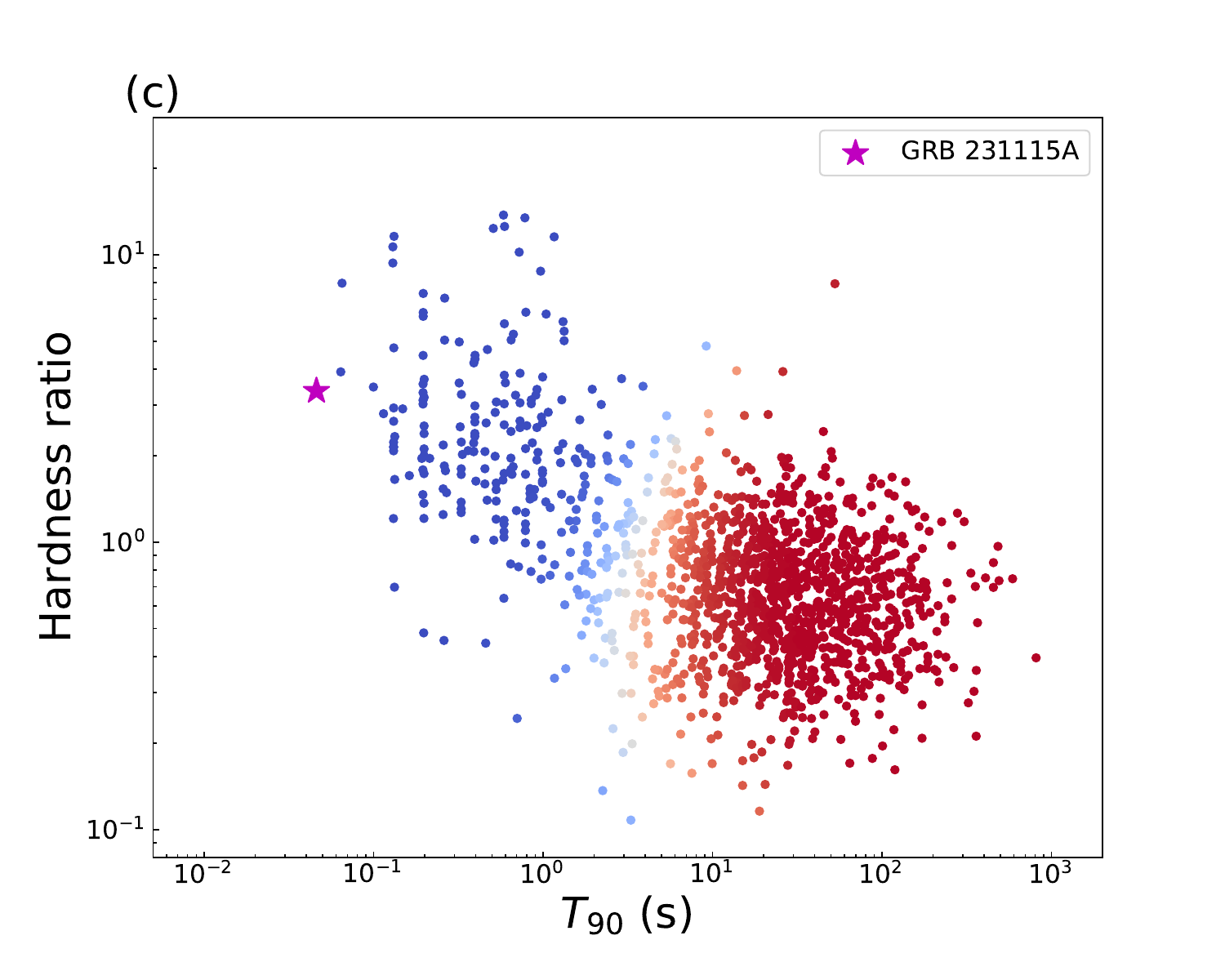}
    \includegraphics[width=0.49\textwidth]{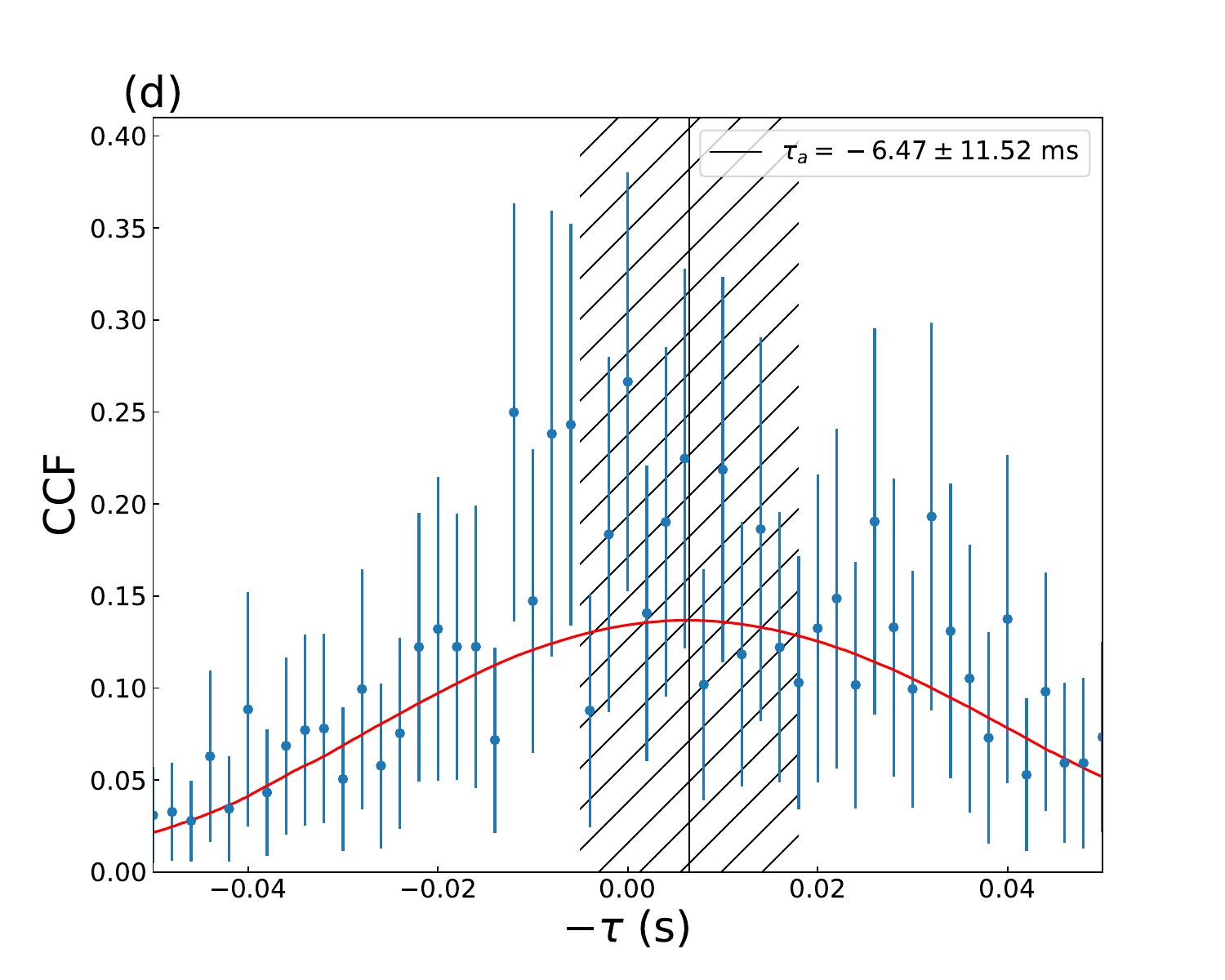}
    \caption{Comparison of GRB 231115A characteristics with other GRBs. (a) depicts the $E_{\rm p,z}$ - $E_{\gamma,\rm iso}$ diagram. The light blue and light red points represent the data of Type I and Type II GRBs with known redshifts, respectively, while the corresponding dashed lines delineate the 2$\sigma_{\rm cor}$ correlation regions \citep{minaev2020p,2020AstL...46..573M}. The black trajectory illustrates the redshift calculation ranging from 0.0001 to 5. Additionally, the red points, black squares, and blue triangles denote the results of GRB 231115A at distances of 3.5 Mpc, redshifts of 0.1, and 1, respectively. (b) presents the $T_{90}$-\textit{EH} relation, with the same colors and labels as in (a). (c) presents the $T_{90}$ and hardness ratio of known GRBs \citep{goldstein2017ordinary},  with the purple star representing GRB 231115A. (d) presents the result of spectral lags, with the error derived from the Monte Carlo simulation of the light curve. \citep{ukwatta2010spectral}.} 
    \label{fig:chara}
\end{figure}

\subsection{$E_{\rm p,z}$ - $E_{\gamma,\rm iso}$ and EH-$T_{90,z}$ relation}\label{sec:amati}
Through the use of the fitting result of the spectra analysis in Section \ref{sec:spec_ana} We intended to place GRB 231115A on the $E_{\rm p,z}$ - $E_{\gamma,\rm iso}$ relation\citep{amati2002intrinsic}, where $E_{\rm p,z} = (1+z)E_{\rm p}$ is the rest frame peak energy, $E_{\gamma,\rm iso}$ is the isotropic bolometric emission energy, written as 
\begin{equation}
    E_{\gamma,\text{iso}} = \frac{4 \pi d_L^2 k S_{\gamma}}{1+z},
    \label{eq:E_gamma_iso}
\end{equation}
where $d_L$ is the luminosity distance, $S_{\gamma}$ is the energy fluence in the gamma-ray band, and $k$ is the correction factor, which can correct the energy range of the observer frame to the energy range of 1--10,000 keV in the rest frame. The correction factor $k$ \citep{bloom2001prompt} writes as
\begin{equation}
    k = \frac{\int_{1/(1+z)}^{10^4/(1+z)}  E N(E) {\rm d} E }{\int_{e_1}^{e_2} EN(E) {\rm d} E},
    \label{eq:k_cor}
\end{equation}
where $e_1$ and $e_2$ correspond to the energy range of the detector. In our calculations, the cosmological parameters of \emph{H$_{0}$} = $\rm 69.6 ~kms^{-1}~Mpc^{-1}$, $\Omega_{\rm m}= 0.29$, and $\Omega_{\rm \Lambda}= 0.71$. 
As shown in of Figure~\ref{fig:chara}(a), the solid black line 
represents the calculation result of redshift from 0.0001 to 5. 
Obviously, when the luminosity distance of GRB 231115A is 3.5 Mpc (that is, located in the M82 galaxy), it falls within the region of GFs. When it falls within the short GRB region, its redshift is 0.1, marked as a black square on the figure. In addition, we also calculated the results when taking the typical redshift of short GRB ($z=1$), marked as a blue triangle. Additionally, \cite{minaev2020p} proposed a new classification scheme combining the correlation of $E_{\gamma, {\rm iso}}$ and $E_{p,z}$ and the bimodal distribution of $T_{90}$.
This characteristic parameter $EH$ is written as
\begin{equation}\label{key}
    EH = \dfrac{(E_{p,z}/100{\rm keV})}{(E_{\gamma,iso}/10^{52}{\rm erg})^{0.4}}.
\end{equation}
As shown in Figure~\ref{fig:chara}(b), the solid black line also represents the calculation results of the same redshift in $E_{{\rm p},z}$--$E_{\gamma,\rm iso}$ diagram. As the luminosity distance increases, GRB 231115A moves from the region of GFs to the region of short GRBs.
And a summary of GRB 231115A and other known GFs is shown in Table \ref{tab:tab2}.

\setlength{\tabcolsep}{1mm}{}
\begin{deluxetable}{c|cccc|cccccccccccccc}
\label{tab:tab2}
\tabletypesize{\small}
\tablecaption{Summary of characteristics of know GFs}
\tablehead{\colhead{\bf GF event date} &\colhead{790305\tablenotemark{a}}& {960618\tablenotemark{b}} &\colhead{980827\tablenotemark{c}} &\colhead{041227\tablenotemark{d}}  &\colhead{051103\tablenotemark{e}} &
070201\tablenotemark{f} &\colhead{070222\tablenotemark{g}}&\colhead{200415\tablenotemark{h}} &\colhead{231115\tablenotemark{i}}}
\startdata
\hline
{\bf Origin}&&&&&&&&\\
SGR or Host galaxy& SGR 0526-66& SGR 1627-41& SGR 1900+14& SGR 1806-20& M81 &M31 & M83 & NGC 253 &M82 \\
Luminosity Distance (Mpc)& 0.050& 0.011& 0.013& 0.0087& 3.6& 0.74& 4.5& 3.5 & 3.5 \\
\hline
{\bf Prompt characteristics}&&&&&&&&\\
Duration (s)& $\lesssim 0.25$& $\sim 0.6$& $<$1.0& $<0.2$& 0.17& 0.01& 0.038& 0.1 & 0.046 \\ 
$E_p$ (keV)& $\sim 500$& $\sim 150$& $>250$& $\sim 850$& $\sim 900$& $\sim 300$& $\sim 1300$& $\sim 1100$ & $\sim 600$ \\
$E_{\gamma,\rm iso}$ ($10^{45}$ erg)&  0.7 & 0.01& 0.43& 23& 70& 1.5& 6.2& 13& 1.3\\
\enddata
\tablenotetext{a}{\cite{1979Natur.282..587M,Mazets1982,2008ApJ...680..545M}.} 
\tablenotetext{b}{\cite{1999ApJ...519L.151M} and \cite{2001ApJS..137..227A}.}
\tablenotetext{c}{\cite{1999Natur.397...41H}, \cite{1999AstL...25..635M} and \cite{2007ApJ...665L..55T}.}
\tablenotetext{d}{\cite{2005Natur.434.1107P} and \cite{2007AstL...33....1F}.}
\tablenotetext{e}{\cite{2007AstL...33...19F}.}
\tablenotetext{f}{\cite{Mazets2008}.}
\tablenotetext{g}{\cite{2021ApJ...907L..28B}.}
\tablenotetext{h}{\cite{2021Natur.589..211S}.}
\tablenotetext{i}{This work.}
\end{deluxetable}

\subsection{Hardness Ratio}\label{}
The Hardness Ratio (HR) of a GRB is usually expressed as the ratio of photon counts in two fixed energy bands. In addition to the bimodal distribution of $T_{90}$ \citep{1993ApJ...413L.101K}, short GRBs are usually harder than long GRBs \citep{2006A&A...447...23H}.
Here we calculate the ratio of the observed counts in 50--300 keV compared to the counts in the 10--50 keV band in $T_{90}$ ($T_0 + [-0.014,~0.032]~\rm s$).  {The calculated HR = 3.36, compared with the previous statistical data \citep{goldstein2017ordinary}, is shown in Figure~\ref{fig:chara}(c). This value is between that of the GRB 200415A's spike (HR = 4.45) and its weak tail (HR = 2.3) \citep{2020ApJ...899..106Y}. }

\subsection{Spectral Lag}\label{sec:spec_lag}
 The spectral lag of GRB is that the high-energy photons arrive earlier than the low-energy photons\citep{norris2000connection,norris2002implications,norris2005long}.
The time delay in different energy bands can be quantified using the cross-correlation function (CCF), which is widely used in the calculation of the GRB spectral lag \citep{band1997gamma,ukwatta2010spectral}.
In general, long GRB exhibit a relatively significant spectral delay \citep{norris2000connection,gehrels2006new}, but not for short GRBs \citep{norris2006short}. Besides, a fraction of short GRBs even show negative lags \citep{yi2006spectral}. We calculated the CCF of the GRB~231115A time series in the energy band 100-150 keV and 200-250 keV with $T_0 + [-0.050,~0.0150]~\rm s$. 
We estimated the uncertainty of the lag by Monte Carlo simulation \citep{ukwatta2010spectral}. The corresponding spectral lags of GRB~231115A is $\tau_a =  -6.47 \pm 11.52 ~\rm{ms}$ as shown in Figure~\ref{fig:chara} (d). {In the observer frame, the mean values of spectral lags for long and short GRBs are $\tau^L = 102.2 \pm 38.1$ ms and $\tau^S = -0.73 \pm 7.14$ ms, respectively \citep{2015MNRAS.446.1129B}. The tiny spectral lag makes GRB~231115A consistent with short GRBs. Note that the spectral-lag of GRB~200415A is also tiny \citep{2020ApJ...899..106Y}.}

\subsection{Initial Lorentz factor}
For such large energy ($\sim 10^{44} - 10^{46}$ erg) to be released within a very short time scale ($\sim$ 0.1 s), GFs are bound to produce a fireball similar to the classic GRBs. Therefore, we consider the calculation of the initial Lorentz factor of GFs under the framework of fireball photosphere emission. Through the thermal component identified by spectral analysis, we can constrain the Lorentz factor $\Gamma$ of the relativistic outflow \citep{pe2007new}. Assume that the radius of the photosphere is larger than the saturation radius, the Lorentz factor is calculated as
\begin{eqnarray}
    \Gamma=\left[(1.06)(1+z)^{2}d_{L}\frac{Y\sigma_{\rm T}F^{\rm ob}}{2m_{p}c^{3}\Re}\right]^{1/4},
    \label{Lorentz}
\end{eqnarray}
where $d_L$ is the luminosity distance, $\sigma_{\rm T}$ is the Thomson scattering cross section, and $F^{ob}$ is the observed flux.
We set $Y$ = 1 in our calculations, which is the ratio between the total fireball energy and the energy emitted in the gamma rays.
$\Re$ is expressed as $\Re=\left(\dfrac{F^{\rm ob}_{\rm thermal}}{\sigma T^{4}_{\rm max}}\right)^{1/2}$, where $F^{\rm ob}_{\rm thermal}$ is the thermal radiation flux and $\sigma$ is Stefan’s constant.
We calculated the initial Lorentz factors at different luminosity distances (3.5 Mpc, $z = 0.1$, $z = 1$), which are $\Gamma_1=70$, $\Gamma_2=250$, and $\Gamma_3=660$, respectively.
{The initial Lorentz factor of GRB 231115A in the GF scenario is consistent with that of GRB 200415A (23 $<$ $\Gamma_0$ $<$ 196) provided by \cite{2022MNRAS.517.3881L} and derived from the Fermi-LAT data analysis~\citep{2020ApJ...903L..32Z}.}

\section{Multi-wavelength Afterglow emission}\label{sec:mo}
{The interaction between the relativistic outflow of GRB or GF and the circumburst medium (CSM) could drive a forward shock (FS). Shock-accelerated electrons produce multi-wavelength emission via the synchrotron radiation and the inverse Compton scattering. By considering the standard FS model (for more details, see \cite{2024ApJ...962..115R}), we model the afterglow emission of GRB~231115A in two scenarios.}

\subsection{The scenarios of GRB and GF}
{For the FS model ($F_{\rm FS}=F(t,\nu,\epsilon_e,\epsilon_b,f_e,p,n_0,E_{\rm k},\Gamma_0,\theta_j)$) in both scenarios, {we assume the fraction $\epsilon_e$($\epsilon_b$) of the shock energy density is converted into energy of relativistic electrons (magnetic field) is 0.01 (0.001), the fraction of electrons accelerated $f_e$ is 0.05, the electron energy distribution index $p$ is 2.2, and the medium number density $n_0$ is 0.01 ${\rm cm^{-3}}$.}
We use different origin-related parameters in the GRB scenario and GF scenario.
For the GRB scenario with a redshift of 1 (0.1), the half-opening angle $\theta_j$ in radians is 0.1, and the initial Lorentz factor $\Gamma_0=660~(250)$. {We assume that the kinetic energy is equivalent to the energy released in the gamma-ray band, that is, $E_{\rm k} = 4.55\times10^{51}~(2.18 \times 10^{49})~\rm erg$.} For the GF scenario at 3.5 Mpc, the parameters we used are $\Gamma_0 = 70$, $E_{\rm k} = 1.3 \times 10^{45} {\rm~erg}$, and $\theta_j=\pi /2$, respectively.} 

\begin{figure}[!htp]
    \centering
    \includegraphics[width=0.45\textwidth]{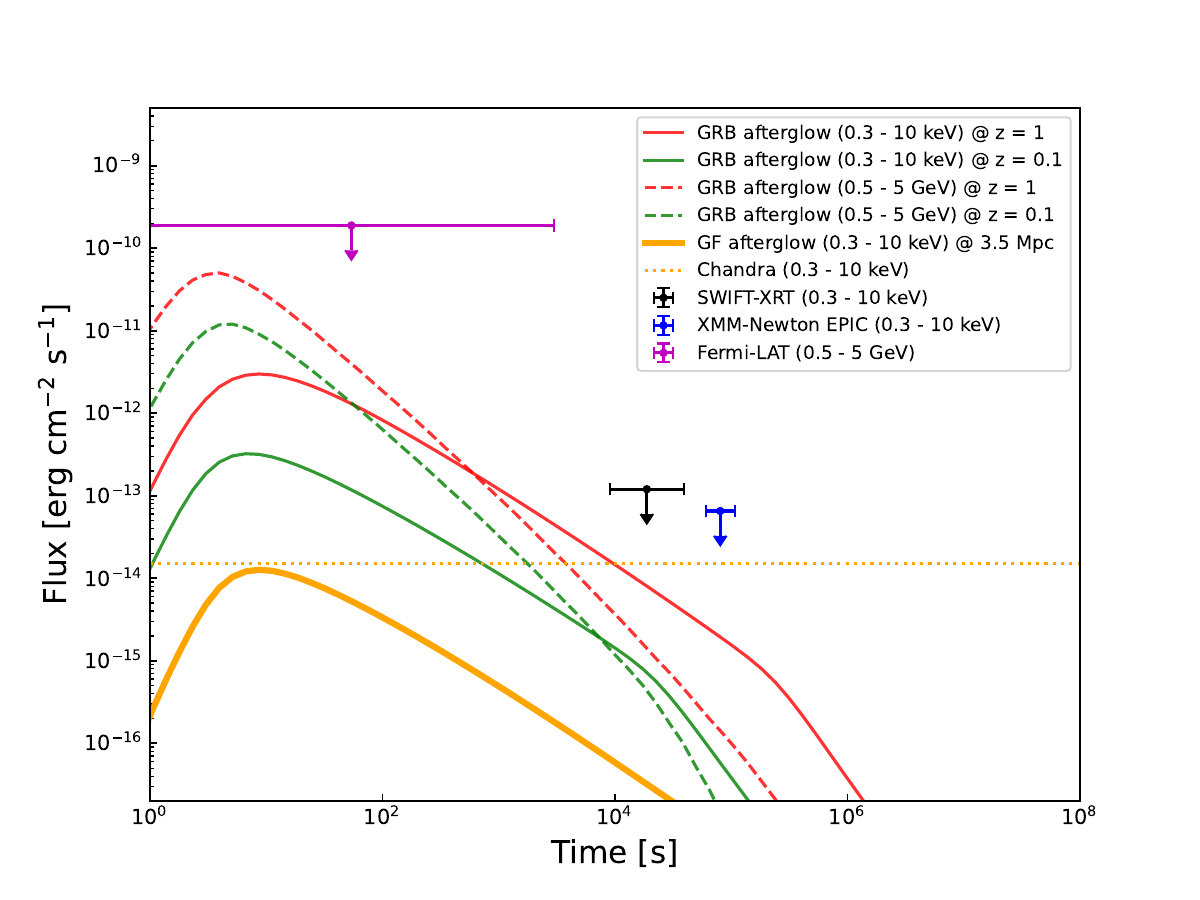}
    \caption{{The light curves of the FS model in different scenarios and the observational upper limits.}} 
    \label{fig:FS_LC}
\end{figure}

{The modeled afterglow light curves in two scenarios are shown in Figure~\ref{fig:FS_LC}. This figure also includes observational upper limits reported by some facilities. In X-ray band (converted to the 0.3--10 keV range), the black and {blue} dots with downward arrow show the upper limit for Swift-XRT \citep[$1.2 \times 10^{-13} \rm~erg~cm^{-2}~s^{-1}$,][]{2023GCN.35064....1O} and XMM-Newton \citep[$6.5 \times 10^{-14} \rm~erg~cm^{-2}~s^{-1}$,][]{2023GCN.35175}, and the yellow dashed line represents the upper limits for Chandra \citep[$1.5 \times 10^{-14} \rm~erg~cm^{-2}~s^{-1}$,][]{2023GCN.35054....1K}. In the gamma-ray band, the magenta dot with a downward arrow represents the Fermi-LAT upper limit ($1.89 \times 10^{-10} \rm~erg~cm^{-2}~s^{-1}$) calculated in Sec.~\ref{sec:fermi-lat}. The modeled results of this set of parameters show that the expected afterglow emission in the GRB scenario does not exceed the observational upper limits of Fermi-LAT, Swift-XRT, and XMM-Newton, but exceeds the observation upper limit of Chandra based on historical exposure. The GF scenario is just below the upper limit of Chandra's observation, providing the tendency of this origin.}

{Based on the current modeling results, we can offer constraints on its progenitor. Following the model proposed by \cite{2016MNRAS.461.1498M}, and considering that the magnetar generating GRB~231115A is surrounded by a wind bubble, the isotropic outflow associated with GF interacts with the nebular, generating afterglow emission. The afterglow emission is associated with the nebular density $n_{\rm nb}$, which is estimated to be}
\begin{equation}
n_{\mathrm{nb}} \simeq 2.1 \times 10^3 \mathrm{~cm}^{-3} M_{\mathrm{nb},-7} P_{i,-0.5}^{6 / 5} M_{\mathrm{ext}, 0.5}^{3 / 5} V_{\mathrm{ext}, 8.75}^{-9 / 5} T_{10\;\mathrm{yr}}^{-3},
\end{equation}
{where $M_{\rm nb} \sim 10^{-7} ~M_{\odot}$ is the nebular mass (which is comparable to the ejecta mass for 2004 GF of SGR~1806-20~\citep{Gaensler_2005Natur.434.1104G}, $P_i \sim 300 {\rm~ms}$ is the initial rotation period of the magnetar (which is the typical value for Galactic pulsars~\citep{FG_2006ApJ...643..332F}), $M_{\mathrm{ext}} \sim 3 ~M_{\odot}$ is the mass of supernova (SN) ejecta, $V_{\mathrm{ext}} \sim 5000 ~\rm km~s^{-1}$ is the velocity of SN ejecta (which is the typical value for super-luminous SN \citep{2016ApJ...818...94K}). It is noted that for the younger magnetar, the density of the nebula becomes larger, {leading} to a stronger afterglow emission. Combined with Chandra's observation, we find that the age of the magnetar should be larger than $\sim 560 {\rm~yr}$, which is just smaller than the upper limit.}

\begin{figure}[!htp]
    \centering
    \includegraphics[width=0.32\textwidth]{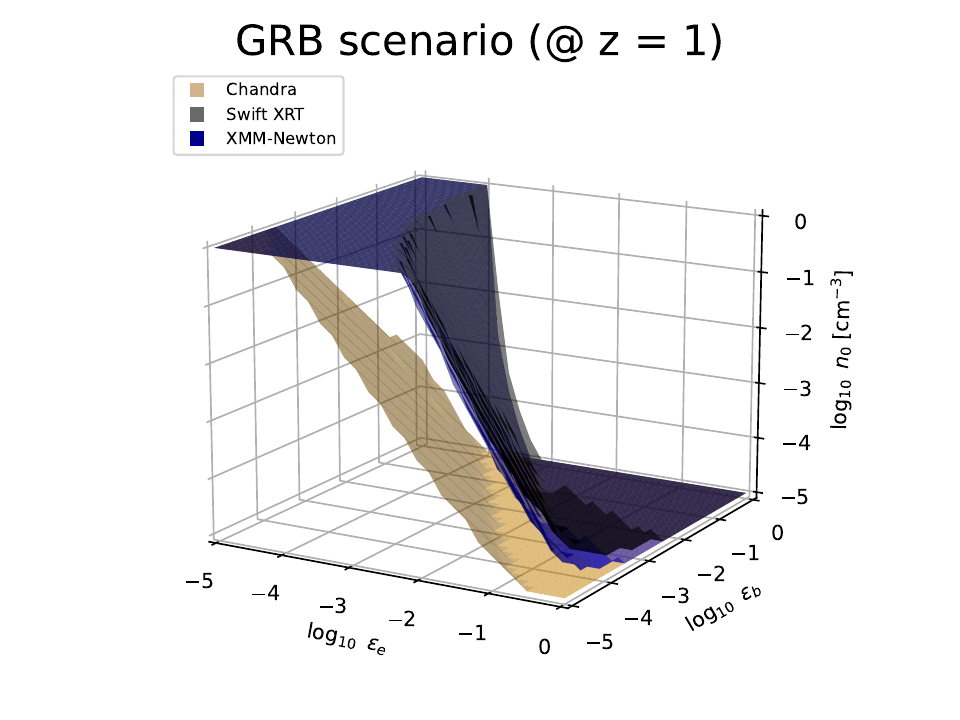}
    \includegraphics[width=0.32\textwidth]{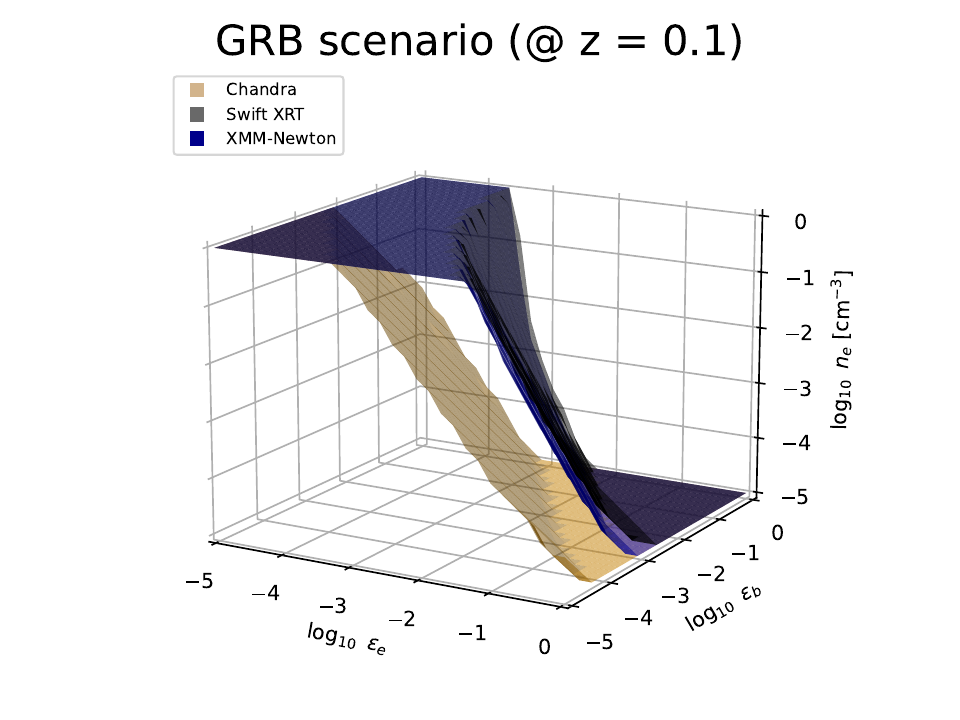}
    \includegraphics[width=0.32\textwidth]{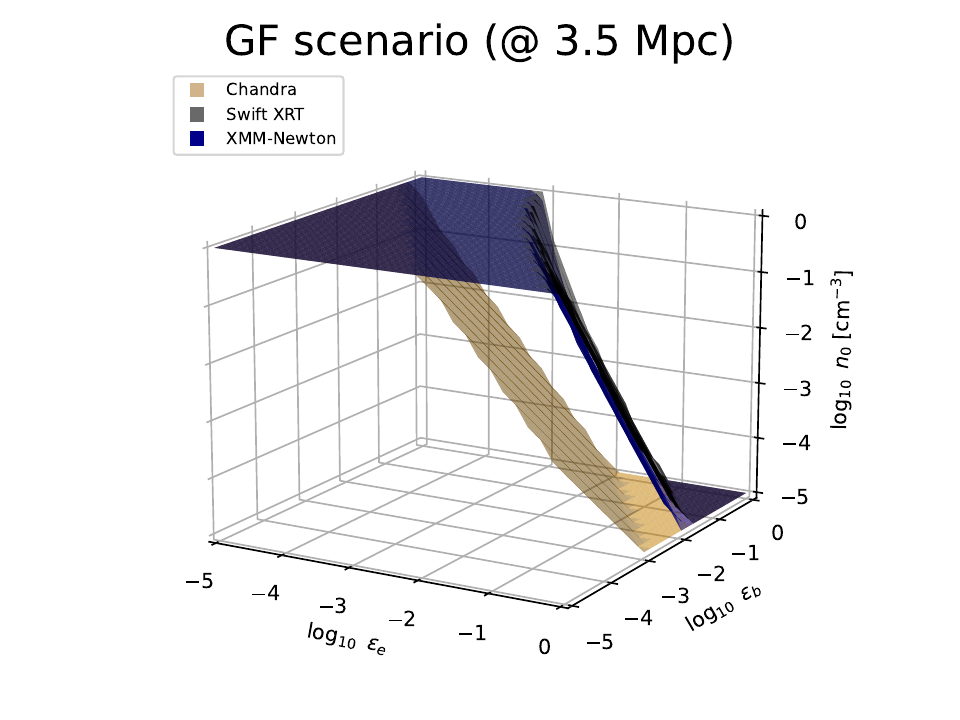}
    \caption{{The observation upper limit corresponds to the parameter space {searching} of the FS model.}}
    \label{fig:para_space}
\end{figure}

{In order to further probe the parameter space of the modeling, we search for $\epsilon_e$, $\epsilon_b$ and $n_0$ from -5 to 0 in logarithmic space for different scenarios in X-ray (0.3--10~keV), as shown in Figure \ref{fig:para_space}. The orange, gray, and blue surfaces represent the upper limits of observations by Chandra, Swift-XRT, and XMM-Neweton under different parameter settings, respectively. The allowed parameters or search boundaries are below the colored surface. It is evident that the GF scenario allows for a larger parameter space than the GRB scenario.}

\subsection{Multi-wavelength observation potential of future GF}
\begin{figure}[!htp]
    \centering
    \includegraphics[width=0.45\textwidth]{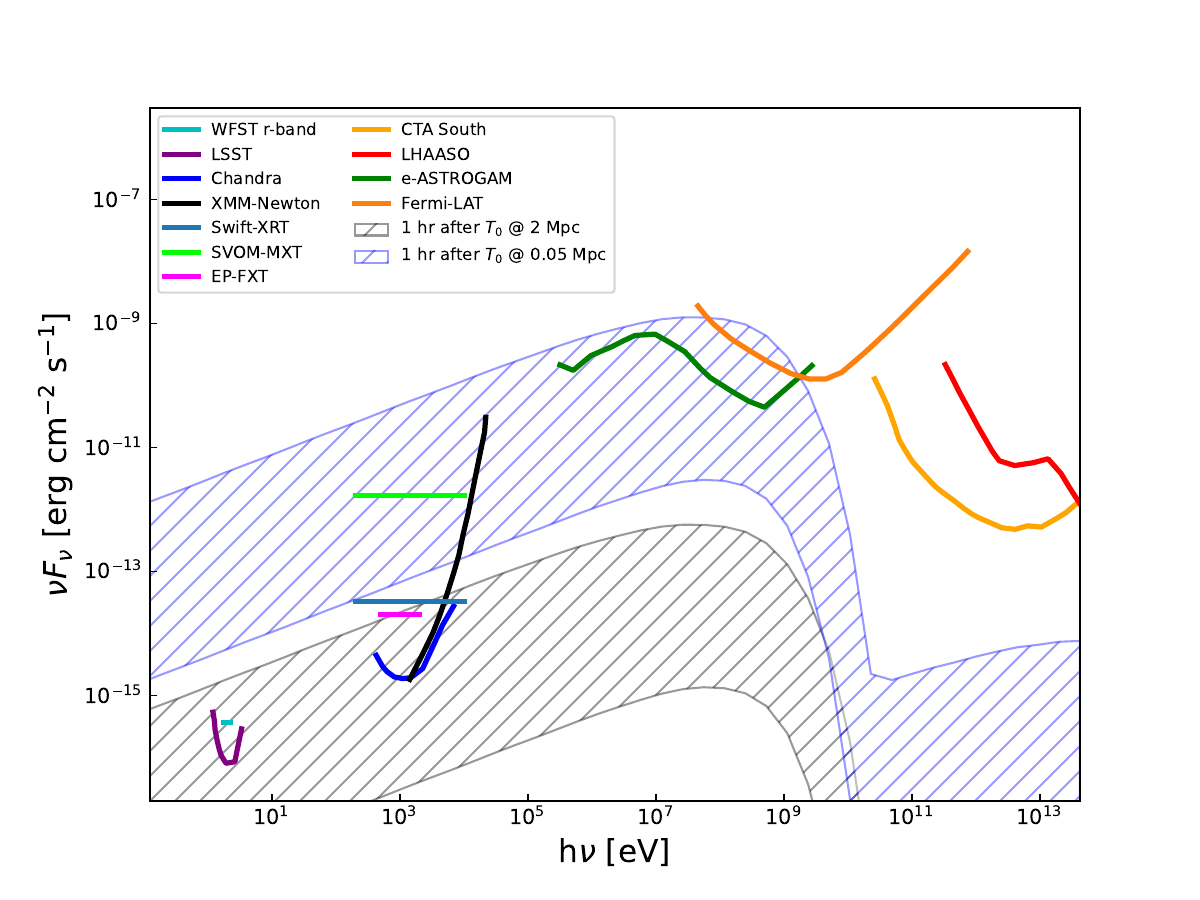}
    \caption{{The SED predicted by the FS radiation of the  GF outflow and the detection sensitivity of some observation facilities. The exposure time $T_{\rm exp}$ for all sensitivity curves {we} used is 1 hour. The sensitivity curves for Fermi-LAT (orange line), e-ASTROGAM (green line), LHAASO (red line), and CTA South (yellow line) are derived from \cite{2018JHEAp..19....1D}. The sensitivity curves for Chandra (blue line) and XMM-Newton (black line) are from \cite{Lucchetta_2022JCAP}. Other sensitivity curves include: the r-band of the Legacy Survey of Space and Time (LSST; purple line) from \cite{Yuan_2021ApJ...911L..15Y}, the r-band of the Wide Field Survey Telescope (WFST; cyan line) from \cite{2023RAA....23c5013L}, Swift-XRT (darkcyan line) from \cite{2005SSRv..120..165B}, SVOM-MXT (lime line) from \cite{2015arXiv150700204G}, and the Einstein Probe (EP; magenta line) from \cite{Yuan_2022hxga.book...86Y}.}}

    \label{fig:SED_FS}
\end{figure}

{We can roughly estimate the triggering of GF by existing gamma-ray monitors. For example, considering the current detection capability of Fermi-GBM (0.71 ph/cm$^2$/s$^{-1}$ in 50--300 keV)\footnote{\url{https://gammaray.nsstc.nasa.gov/gbm/instrument/}}, if {a GF is} to release $E_{\gamma,\rm iso} > 10^{45}$ erg of energy within 0.1 s (assuming a blackbody spectrum with $kT$=100 keV), it {will} be expected to be detected within approximately 10 Mpc. As shown in the blue (grey) {shaded area} in Figure \ref{fig:SED_FS}, we {model} the spectral energy distribution (SED) starting 1 hour after trigger (with 1 hour exposure time) in the parameter space {including} $E_{\rm k}$ from 10$^{45}$ to 10$^{47}$~erg and $\Gamma_0$ from 10 to 90, {with} $\epsilon_e = 0.01$, $\epsilon_b = 0.001$, $f_e = 0.05$, $n_0 = 0.01 {\rm~cm^{-3}}$, $p = 2.2$ and $d_L$ = 0.05(2)~Mpc. The sensitivity of the relevant facilities is converted to an exposure time of 1 hour by assuming $F_{\rm limit} \propto T_{\rm exp}^{-0.5}$.}

{For the grey shaded area ($d_L = 2$ Mpc), the afterglow emission associated with GFs can be detected by optical telescopes (LSST and WFST), as well as X-ray telescopes (Chandra, XMM-Newton, and EP-FXT). For the blue shade area ($d_L = 0.05$~Mpc), in addition to optical and X-ray bands, the afterglow emission at the gamma-ray band also has the opportunity to be detected. 
Note that here we choose relatively low microphysical parameters corresponding to a lower flux, and the afterglow with higher values of microphysical parameters are more likely to be observed. According to the modeling results and the discussion in \cite{2005MNRAS.361..965F}, we are also expected to observe possible high-energy (Synchrotron Self-Compton) emission of afterglow for such events in our own galaxy. Therefore, timely follow-up observations are crucial.}

\section{Summary and Discussion}\label{sec:sd}
{In this work, we conducted a detailed analysis of the prompt emission of GRB 231115A based on Fermi observation data and modeled its afterglow emission under different scenarios. By combining the observations with our analysis results, we can list the characteristics of this event, which may originate from a magnetar GF, as follows:}
\begin{itemize}
\item{{For GRB 231115A detected by Fermi-GBM, the lack of detection of electromagnetic counterparts results in the highest position accuracy being 2 arc minutes, as provided by INTEGRAL \citep{2023GCN.35037....1M}. The error range of the burst position almost entirely covers the M82 galaxy, suggesting a high probability (coincidence probability $\sim$ 0.03\%) of the event originating from this galaxy.}}
\item{{In the Fermi-GBM data analysis of the prompt emission of GRB 231115A, we found that its duration ($\sim$ 46 ms) is exceptionally short, and no significant quasi-periodic signal was detected. The empirical relationships ($E_{\rm p,z}$ - $E_{\gamma,\rm iso}$, EH-$T_{90,z}$) at different luminosity distances, as well as some characteristics (HR, spectral lag), are consistent with short GRBs and GFs, and it exhibits a quasi-thermal spectrum. Considering the photospheric emission at various luminosity distances (3.5 Mpc, $z = 0.1$, $z = 1$), we can roughly estimate the initial Lorentz factors as $\Gamma_1=70$, $\Gamma_2=250$, and $\Gamma_3=660$ respectively.}}
\item{{Based on the FS model calculations and the observational upper limits from Swift-XRT, XMM-Newton, and Chandra, we found that the parameter space required in the GRB scenario is more challenging than that of the GF scenario.}}
\end{itemize}

{The light curve of the prompt emission for GRB 231115A exhibits a slight difference from the GF case of GRB 200415A, showing a slower rise, as depicted in the top panel of Figure \ref{fig:LC_PAR}. A similar GF (i.e., 960618) occurred in SGR 1627–41 \citep{1999ApJ...519L.151M,1999ApJ...519L.139W}, which also displayed a relatively slow rise. These two events (GRBs 960618 and 231115A) may represent another subspecies of GFs and deserve further investigation. Since the location of this event is associated with the nearby M82 galaxy, and its characteristics are basically consistent with those of GRB 200415A, the origin of the magnetar GF has become a consensus in recent work and reports.}

{Due to the lack of observations of multi-wavelength electromagnetic counterparts of such events, our conclusions suffer from some uncertainties. For example, in the short GRB scenario, the non-detection of the afterglow emission may be due to the very small $\epsilon_{\rm e}$ and/or $\epsilon_{\rm B}$, as inferred in for instance GRB 170817A \citep{2023ApJ...943...13W}. In the GF scenario, the allowed parameter space is wider. Interestingly, if GRB 231115A was indeed associated with M82, then the GF scenario would favor over the short GRB scenario by a factor of $\sim 180,000$  \citep{2021ApJ...907L..28B,2023GCN.35038....1B}. 
In the plausible GF scenario, the age of the magnetar is estimated to be $\sim 560$ years. This age however depends on some further assumptions on the afterglow parameters such as $\epsilon_e$, $\epsilon_b$, $f_e$, and $n_0$. For proper parameters, the multi-wavelength afterglow radiation of the GF outflow is detectable for some current and upcoming facilities (see Figure \ref{fig:SED_FS}).
}

{In the future, with a quick analysis of gamma-ray monitor data (e.g., Fermi-GBM, SVOM-GRM), we can evaluate whether the trigger is a GF candidate and roughly estimate the observational potential of relevant facilities (e.g., WFST, EP-FXT). Benefiting from the observation mode of the upcoming SVOM, the equipped VT can automatically conduct quick follow-up observations. Rapid follow-up observations are also expected to enable direct measurement of the redshift. This is expected to advance research on their localization and reveal their progenitors, like the property of the magnetar \citep{2023MNRAS.524.6004W}. The magnetar can be associated with different types of bursts \citep[e.g.,][for a review]{Kaspi_2017ARA&A..55..261K}, including GFs and the fast radio bursts (FRBs). Two types of models for the origin of fast radio bursts associated with magnetars have been debated~\citep[e.g.,][for a review]{2023RvMP...95c5005Z}, and the multi-wavelength observations of GFs may also help reveal the possible origin of FRBs.}

\section*{Acknowledgments}
W.Y. Thanks to Lu-Yao Jiang, Lei Lei, Prof.~Da-Ming Wei, and Prof.~Yi-Zhong Fan for helpful discussions. We acknowledge the use of the Fermi archive's public data. This work is supported by the Natural Science Foundation of China (NSFC) under grants of No. 11921003, No. 11933010, No. 12003069, No. 12321003 and No. 12225305, and the Strategic Priority Research Program of the Chinese Academy of Sciences, Grant No. XDB0550400.

\vspace{5mm}

\bibliography{sample631}{}
\bibliographystyle{aasjournal}

\end{document}